\documentstyle[prl,aps]{revtex}

\newcommand{\eq}[1]{\begin{equation}\label{#1}}
\newcommand{\eqs}{\begin{equation}}
\newcommand{\en}{\end{equation}}
\begin{document}

\twocolumn[{

\title{High dimensional properties of quenched noise growth models.}
\author{Omri Gat and Zeev Olami}
\address{Chemical Physics Department, Weizmann Institute of Science \\
Rehovot, Israel 76100\\\today}
\maketitle
\widetext
 \leftskip 54.8pt
 \rightskip 54.8pt
\begin{abstract}
We discuss the behavior of bounded slope quenched noise invasion models in 
high dimensions. 
We first observe that the roughness of such a steady state interface is generated
by the combination of  the roughness of the invasion process $\chi_c$ 
and the roughness of the underlying interface dynamics. In high enough dimension  
we argue that $\chi_c$ decreases to zero. This defines a critical
dimension for the problem, over which it reduces to the correlated annealed 
dynamics, which we show to have the same roughness as the annealed
equation at five dimensions.
We argue that on the Cayley tree with one additional height coordinate 
the associated processes are fractal. The critical behavior is anomalous
due to strong effects of rare events.
Numerical simulations of the model on a Cayley tree and high dimensional lattices
support those theoretical predictions.
\end{abstract}
\vskip 5mm
}]
\narrowtext
Kinetic roughening of interfaces during nonequilibrium growth has attracted  a
lot of attraction lately. Initially  the focus of the theory  was on
interfaces which 
are driven by  random noise which does not depend on the position of
the interface (annealed noise)\cite{gen}. A challenging problem
motivated by experiments is what is the effect of disorder  in the
growth medium (quenched disorder). Examples of systems with such
disorder are invasion of liquid in porous media and magnetic domain
wall growth. It was discovered experimentally that growth in systems with
quenched disorder generates exponents which are different significantly
from the known annealed exponents.\cite{gen}

In growth  problems there is usually a marked distinction between the 
behavior of the quenched noise problem and the annealed versions,
and they define different universality classes. In particular the
roughness exponent $\chi$ of the quenched growth is different from
$\chi_a$, the roughness of the annealed model.    
The roughness exponent, $\chi$, is defined by
$W=L^\chi$, where $W$ is the system width and $L$ is the size of the
system. An important example is bounded slope growth, which may be
described by the KPZ equation \cite {kpz} for the annealed case~\cite
{gen}. The quenched model \cite{gen,dis,snne,opz2} has a
different roughness. Growth in quenched
disorder is characterized by the
existence of a critical force level. When interfaces are
driven at this critical force, critical behavior appears,
in the sense that interfaces become self-affine, and have infinite
correlation length, associated with processes of unbounded size. 

In a set of recent papers\cite{opz2,opz1} we pointed out that the main distinction
between growth in annealed disorder and growth in quenched randomness is
that the growth in quenched randomness can be decomposed into consecutive
associated processes (APs), which are defined as a series of activations
below a certain threshold value $f_0$ of the quenched noise field. These
APs are connected sets of sites which are delimited
by a network of blocking surfaces.
The APs have an accompanying roughness exponent $\chi_c$ which is defined by the 
scaling of the height of the APs with respect to their radius. It is
usually assumed that $\chi_c$ defines the 
roughness of the interface\cite{dis,high-d}; actually, it is only 
a lower bound for the overall roughness of
the surface, not necessarily identical with it\cite{opz2}.

In this study we concentrate on discrete bounded slope models\cite {dis,snne} which
are believed to belong to the quenched KPZ universality
class~\cite{opz2,amrl}. In these models the slope of the growing surface
is constrained to be below a threshold value.
  We  discuss the
behavior of those models in high dimensions. We show that
$\chi_c$ should go to zero in high enough 
dimension. We argue that the APs should become fractal
objects, with a fractal dimension which remains bounded in all dimensions. To
test this we present numerical analysis of systems up to six
dimensions and on a Cayley tree with a height. We find that there is a
nontrivial transition between the high dimensional behavior and the
Cayley tree. We discuss the relationship between our conclusions and a
recent analysis\cite {high-d}.

Analyzing the overall interface growth, the distinction between
the annealed and the quenched problems becomes less important
as soon as the APs become flat (or fractal). The
quenched problem then becomes equivalent to the correlated annealed noise
problem.
We analyze the relevance of such noise. We argue that the
roughness exponents of this problem are the same as the annealed KPZ
exponents at 5 dimensions and above.

The discrete Buldyrev-Sneppen model \cite{dis,snne}, which belongs to the
quenched KPZ universality class,  may be defined on any
space of the form ${\cal S}={\cal L}\times{\bf Z}^+$, where $\cal L$ is an
arbitrary substrate lattice, and ${\bf Z}^+$, the set of positive
integers in the direction of interface growth. $\cal
L$ will be defined as a hypercubic lattice $R^d$ in $d$ dimensions and
as a Cayley tree in the infinite dimensional limit\cite {tho}.  At each
point $s$ of $\cal S$ 
there is a random pinning force $f(s)$ distributed uniformly
between 0 and~1. The interface is a graph over $\cal L$. 
The growth process is as follows: the site of minimal $f$ value is
activated by  increasing the interface height at the point,
followed by growth at nearest neighbor sites, if the local
slope exceeds a threshold value.

The growth of the interface can be represented as a series of associated
processes, defined as sets of activations below a threshold value $f_0$ of the
quenched field  $f$.  An activation above $f_0$ is
concludes  an AP and starts a new one. The
APs of $f_2$ may comprise several $f_1$ APs if $f_1<f_2$. The
largest APs are the $f_c$ APs, where $f_c$ is the critical $f$
value, {\it i.e.}, the largest to be activated in an infinite system.
$\Delta f\equiv f_c-f_0$, measures the criticality of the APs. 
The APs are thus random processes characterized by the number of
activations $s$, and the r.m.s radius of the projection on $\cal L$ of
the process, $r_\parallel$.

The following scaling forms are known to be valid for $d$-dimensional APs:
\eq{sc-def} P(s)=s^{-\tau}g(s/\Delta f^{-\nu}),\quad
P(r_\parallel)=r_\parallel^{-\tau_\parallel}g_\parallel(r_\parallel/\Delta
f^{-\nu_\parallel}),\en
and the fractal dimension $D$ of the APs is defined by
\eq{sc-frc}s\sim r_\parallel^D. \en
The scaling relations
\eq{sc-par} \nu_\parallel=\nu/D\qquad\tau_\parallel-1=D(\tau-1)\en
follow immediately from (\ref{sc-def}) and (\ref{sc-frc}). Another scaling relation 
proved in \cite{opz1} for 1-dimensional APs is
\eq{sc-opz} 1=\nu_\parallel(d+1-\tau_\parallel). \en
This relation reduces the number of independent static exponents to~2.
When $D>d$ the APs are rough, with roughness exponent
$\chi_c=D-d$. It will be shown below that those relations
break down for a Cayley tree.

It is evident that the interface roughness  in this system
cannot be smaller than the value $\chi_c$. This is because from time
to time the surface identifies with a critical surface which has this
minimal roughness\cite {opz2}. As an example one can consider the $1+1$
dimensional case. Since the APs are related to
directed percolation their roughness is $\chi_c =0.63$\cite{dis}. The
roughness for  the annealed dynamics is $\chi_a=0.5$, so that $\chi_c$
dominates $\chi_a$.
On the other hand we should note that the roughness of the annealed
KPZ equation is a lower bound on the roughness of this system. If
$\chi_c=0$, the roughness of the system will be defined by a process
which will be similar to a correlated annealed KPZ dynamics. In other words:
\eq{bounds}
\chi \geq \max(\chi_{a},\chi_c).
\en
The roughness will be generally defined by an interplay between
$\chi_{a}$, $\chi_c$ and the correlations in the system, and can
actually be larger than both $\chi$'s. It has been
already observed that even in the simplest $1+1$ dimensional case the
system roughness seems to be higher then $\chi_c$ \cite{opz2}.  For
higher dimensions the effects should be much stronger since $\chi_c$
decreases. Therefore, we now focus our attention on the dependence of
$\chi_c$ on the lattice dimension $d$.
We note that the usual relationship
$\chi=\nu_\bot/\nu_\parallel$\cite{gen,dis,opz1,high-d,stanley}
is wrong in the general case, this relation
applies only to $chi_c$. Furthermore if $\chi > \chi_c$  
processes initiated from a flat surface such as analyzed in
\cite{high-d,stanley} will be different from the
associated processes we defined here\cite{us}
in scaling properties and in statistics.   

We first present an argument to show that $\chi_c$ must decrease to 0
on the Cayley tree. Consider a large cluster of radius $>r$.
We want to compare the number of activated sites whose distance from a
central site is $r$  to the number of activated sites with a distance
of $r+1$. Since a critical process is stationary, {\it i.e.}, the processes
activated from sites of distance $r+1$ are statistically the same as
those from distance $r$, the ratio of the number of sites on the two
shells should be asymptotically 1 for a critical cluster. Therefore
 the number of sites on the critical cluster grows more slowly
than exponentially, in other words, it is fractal, with a fractal
dimension $D_\infty$. 
Studying the rules of growth  indicates that whenever $\chi_c$ is
positive, the APs cannot be fractal, since rough
clusters have a typical width $r_\bot\sim s^{\chi_c/D}$, which through
avalanches prevents the formation of holes with diameter smaller than $r_\bot$. For
$d\leq d_{c}$ the APs may be multiply connected, but
not fractal. APs with $\chi_c=0$ are objects with order 1 thickness,
which may fractalize.
We can therefore conclude that $\chi_c$ is 0 on the
Cayley tree.

We have shown that $\chi_c$ decreases from $0.63$ to 0 when $d$
increases from 1 to $\infty$. It is very natural to presume that the
decrease of $\chi_c$ as a function of $d$ is monotonic, and this is
indeed verified by numerical simulations but is not critical to the
 following argumentation.
This scenario for the dependence of $\chi_c$ on $d$ motivates a
definition of an upper critical dimension $d_{c}$ where $\chi_c=0$.
$d_{c}$ marks the transition to fractal APs. We can state that
$d_c\leq D_\infty\leq\infty$. In fact, additional arguments, involving
dynamic exponents indicate that $d_c\leq6$ \cite{us}.
Most of the preceding arguments on the behaviour of the AP as a
function of $d$ may be generalized to other growth models such as
quenched Edwards-Wilkinson growth.

Fractal APs in this model will be unusual due to rare events in
which a single site is activated several times, creating large,
compact avalanches which are exceptional in the overall structure
of the AP.  These events, although exponentially
rare, can dominate the tail of the distribution of the APs, and
thus alter the scaling forms (\ref{sc-def}).
We demonstrate this mechanism for the breakdown of the scaling form for $P(s)$ 
on the Cayley tree. Let ${\bar s}(h)$ be the expectation
for the size of an AP given that a single site is of height $h$. For
 $h$ larger then the average height $\bar s$ must grow exponentially,
\eq{s(h)}  {\bar s}(h)\sim\mu^h, \en
with $\mu$ larger or equal to the coordination number of the lattice
minus~1. $\mu$ is an increasing function of $f_0$ for $f_0\leq f_c$.
On the other hand the probability that a site in an $f_0$  AP
reaches a height $h$ should decrease like ${\tilde f}^h$ with $\tilde f$
larger than, but of the order of $f_0$, and also an increasing
function of $f_0$. A rough estimate for the
probability to activate an AP of size $s$ is
\eqs {\sf P}({\bar s}(h)\leq s\leq{\bar s}(h+1))\sim {\tilde f}^{h}; \en
combining with (\ref{s(h)}) we get
\eq{sc-bte} P(s)\sim s^{\log{\tilde f}/\log\mu-1}\equiv s^{\tau(f_0)}. \en
On the Cayley tree, the distribution of
the APs is always a power law, with a non-universal $f_0$-dependent
exponent. This is in accord with the findings of \cite {high-d}.
 This behavior is in marked contrast to the normal
critical behavior (\ref{sc-def}). The difference should be attributed to the
strength of exponentially rare events, which dominate the large $s$
tail of the distribution. 
The main conclusion (\ref{sc-bte}) does not depend on the assumption
of localized distribution. 
A more detailed analysis of bounded slope models on the
Cayley tree will be given elsewhere \cite{us}. 

A similar analysis may be carried out for finite dimensional systems
whenever the APs are flat or fractal, so that an unusually large height
at one site can modify the local structure of an AP.  In contrast to
the Cayley tree, avalanches stemming from such rare events are
expected to grow only as a power of $h$, $s\sim h^{\tilde d}$,
so that the large $s$ tail of $P(s)$ should decay as a stretched
exponential,
\eq{sc-high d} P(s)\sim s^{-\tau}\exp(-as^{1/\tilde d}), \en
instead of a power law as in the Cayley tree. The dependence of $a$ on
 $f_0$ is yet to be investigated.  
Eq. (\ref {sc-bte})  indicates that the mean cluster size does not
diverge near $f_c$. However, our previous scaling relations
(\ref{sc-par}) and (\ref{sc-opz}) and also
simulations in high dimensions indicate that
$\left<s\right>\sim|f_c-f_0|^{-\gamma}$, with $\gamma\geq1$. Hence
 the transition from low dimensional behavior to high dimensional
behavior is anomalous. We conjecture that the scaling form (\ref{sc-bte}) is
obtained as an infinite dimensional limit of (\ref{sc-high d}).

We now turn our attention to the scaling of the growing interface as
a whole.
Above $d_{c}$ the APs are flat, and growth becomes similar to the annealed
process, with the source correlation derived from the distribution and
structure of the flat or fractal APs. In this regime
$\chi$ will remain positive and equal to that of annealed KPZ \cite{gen,kpz}
\eq{an-kpz}\partial_t h({\bf x},t)=\nabla^2 h({\bf x},t)+\lambda 
(\nabla h({\bf x},t))^2+\eta({\bf x},t),\en
where $h$ is the height and the random source $\eta$ has spatial correlations derived
from the statistics of the APs. This motivates the definition of a
second critical dimension $d_i$, such that $\chi_a=\chi$ for 
$d >d_i$.  $d_i$ rather than $d_c$ is the relevant critical dimension
for the overall roughness.

When $d>d_c$,  $\eta$  is Gaussian and, and its
correlation function is derived as follows: Since
the APs are not correlated (for $f_0=f_c$, see \cite{opz1}), 
the noise at two points is correlated if and only if the two sites were
activated in the same AP.  Therefore the source correlation may be
estimated by the probability of such an event, which may be estimated
as
\eq{corr}\begin{array}{r} \left<\eta({\bf x})\eta({\bf y})\right>\sim
P(r_\parallel\geq|{\bf y}-{\bf x}|) |{\bf y}-{\bf x}|^{D-d}\\\sim
|{\bf y}-{\bf x}|^{D(2-\tau)-d}.\end{array} \en

We now analyze the effects of those correlations.
Let $h({\bf k},\omega)$ $=$ $
\int d^d{\bf x}dt\,h({\bf x},t)e^{i{\bf k}\cdot{\bf x}+i\omega t}$ with 
$\eta({\bf k},\omega)$ defined similarly. We also define
the response function
$G({\bf k},\omega)\delta({\bf k}+{\bf k'})\delta(\omega+\omega')$ $=$ $
\left<\delta h({\bf k},\omega)/\delta\eta({\bf k'},\omega')\right>$.
The scaling form of $G$ is $G({\bf k},\omega)=
k^{-z}g(\omega/k^z)$, and the noise correlation is
$\left<\eta({\bf k},\omega)\eta({\bf k'},\omega')\right>=k^\alpha
\delta({\bf k}+{\bf k'})\delta(\omega+\omega')$, where $\alpha=D(\tau-2)$.
We define the ``bare'' height field as
\eqs h_0({\bf k},\omega)=G({\bf k},\omega)\eta({\bf k},\omega)  \en
If the roughness of $h_0$ dominates the annealed white noise 
roughness $\chi_a$,
the system roughness $\chi$ must be larger than $\chi_a$.
The bare correlation is
\eq{bare} \begin{array}{r}\left<h_0({\bf k},\omega)h_0({\bf k'},\omega')\right>=
G({\bf k},\omega)G({\bf k'},\omega')\left<\eta({\bf k},\omega)\eta({\bf k'},\omega')
\right>\\\sim
k^{-2z+\alpha}g(\omega/k^z)^2\delta({\bf k}+{\bf k'})\delta(\omega+\omega'),
\end{array}\en
and the simultaneous correlation is given by
\eq{baresim} k^{-2z+\alpha}\int d\omega g(\omega/k^z)^2\sim k^{-z+\alpha}.\en
From the correlation in (\ref{baresim}) we get an effective bare roughness
$2\chi_0=\max(z-\alpha-d,0)$.

The condition for irrelevance becomes
\eq{dom-sc} 2\chi>z-\alpha-d. \en
(\ref{dom-sc}) is always satisfied for uncorrelated noise
($\alpha=0$), and also reproduces the known condition for irrelevance
of the noise correlation in 1 dimension, $\alpha>-1/2$ \cite{gen}.
Numerical simulations give $d_c\sim6$ (see below), and
values for $D$ and $\tau$ given in table~1.  
Inequality (\ref{dom-sc}) is satisfied for those values, so
$d_i\leq d_c$. 


We performed numerical simulations of the Buldyrev-Sneppen model in order to
examine  the previous theoretical scenario. The simulations consisted of 
growth in a hypercubic lattice of dimension 1 to 6, and
growth on a Cayley tree as an effective infinite dimensional
lattice. The growth on finite-dimensional lattices was simulated using the
constant current algorithm starting from a flat surface. The search
for the global minimum which is required in this approach was coded in
an efficient manner, so that about $10^{10}$ activations were simulated
in each case. The main computational constraint, however, was memory
area, which dictated lattices of small sides, down to 16 in 6
dimensions, where there
was a small but observable scaling range. The simulations were
performed with periodic boundary conditions.

The exponents $\tau_\parallel$,
$\gamma$ and $D-d$ were measured. They are presented in
table~1. These exponents are not independent, and
$\gamma$, for example, is obtainable from the other two using
relations (\ref{sc-def},\ref{sc-opz}). The calculated values of
$\gamma$ 
 agree with the measured values of $\gamma$, thus providing a
consistency check on the simulations. (A possible source of deviations
is a weak size dependence of the exponents.)

It is evident from table~1 that
the APs become flat
 at 6 dimensions, so  $d_c\sim 6$. The
measured $\chi_c$ becomes equal to numerical measurements of $\chi_a$
\cite{kim-kos} at 4 dimensions within numerical accuracy.
Since $\chi_c\ll1$ at 5 dimensions, it is probably
safe to use the criterion (\ref{dom-sc}) for relevance of noise
correlation for KPZ roughness. The criterion is satisfied so
that $4\leq d_i\leq5$. Independent measurements of the overall interface 
roughness\cite{stanley} indicate that this is indeed the case, see table 1.
There is no numerical indication of an upper critical dimension above
which the exponents don't change.

Simulations were performed also of the Cayley tree
version of the model as a limit of high dimensional systems.
Since  clusters are expected
to be fractal, of a volume growing as $\ell^D$ with the cluster
length $\ell$, it is very inefficient to simulate the process on the full
Cayley tree. Therefore we simulated the processes
on a background of a flat blocked surface punctured at
one point. As discussed above these processes might have different
 scaling than APs on the
Cayley Tree because of the strong dependence of the distributions on height
 fluctuations \cite {us,Zeev} but the general predictions still apply. 
The clusters were grown with a constant driving force, until 
either the surface became blocked, or the size of the cluster exceeded
a certain memory limit (about $2\times10^6$ sites). 

To examine  the distribution of AP sizes we simulated $10^9$ clusters
for  each value of the driving force
$f_0$, on a Cayley tree with coordination number 3. As expected, we find that
the transition is anomalous in the sense that the probability
distribution function is a power law for
all values of $f_0\leq f_c$, of the form (\ref{sc-bte}), with no apparent
cutoff. The value of $f_c$ was calculated by
estimating the number of infinite avalanches initiated by the process.
An equivalent definition is that $\tau(f_c)=2$. It is found that
$f_c=.141\pm.001$. The measured value of $\tau$ decreases with $f_0$
and can be fitted with $\tau(f_0)-2\sim(f_c-f_0)^{1/2}$. This is the
same dependence  as 
in the different Cayley tree realization used in \cite {high-d}. The reason for
this seems to be that the APs are flat so that the differences between
the models become irrelevant. The relationship between the models will be discussed
elsewhere\cite{us}. 
The measurement of $D$ is more difficult, but a significant dependence
on $f_0$ could be observed. The measured value at $f_c$ was
$D=6.5\pm0.5$. This translates to twice the value in high-dimensional
hyper-cubic lattices\cite{bunde}.

In this paper we reached an understanding of the structure of quenched noise
growth as a function of dimensionality. We identified the role of the
underlying KPZ dynamics in determining the roughness of the growing interface.
The concept of AP introduced in Ref \cite{opz1} remains useful in this
analysis. The main idea is that those processes possess an intrinsic roughness
$\chi_c$ and that they interact via annealed KPZ dynamics. The system roughness
$\chi$ is a result of this interplay.
We show that $\chi_c$ decreases with dimension to 0 at $d=d_c$,
and moreover the fractal dimension of the APs saturates
to a finite value $D_\infty$. Hence, above $d_c$ the random sequence of APs
is  equivalent
to correlated noise acting on annealed KPZ equation. Analyzing this effect
 we show that above 4~dimensions
$\chi=\chi_a$. Those general considerations apply to any of the various models
for quenched disorder discussed in the literature\cite {gen,us}.
Finally, we have shown that the critical behavior of
the APs is anomalous on the Cayley tree, and made some conjectures as
to the transition to this behavior from finite dimensions.

We thank S. Alexander, I. Procaccia and S. Havlin for useful
discussions. O. G. thanks the Basic Research Fund of the Israeli
Academy of Sciences.  


\begin{table}
\begin{tabular}{|c|c|c|c|c|c|c|}
$d$ & $f_c$    & $\gamma$  & $D-d$ & $\tau_\parallel$ &
$\gamma_{\rm calc}$ & $\chi$\cite{opz2,stanley}    \\\hline
1  & .4610(2)  &   2.03(3) &  .64(2)  &   1.42(2)  &      2.09   &  0.67  \\  
2  & .2002(2)  &   1.53(2) &  .52(2)  &   2.11(2)  &      1.58   &  0.50  \\  
3  & .1148(3)  &   1.29(2) &  .40(2)  &   2.86(2)  &      1.34   &  0.38  \\    
4  & .0785(5)  &   1.21(1) &  .24(3)  &   3.58(2)  &      1.17   &  0.27  \\    
5  & .0583(2)  &   1.13(2) &  .07(3)  &   4.22(3)  &      1.03   &  0.25  \\   
6  & .0445(5)  &   1.08(2) &  .00(5)  &   5.05(5)  &      1      &  0.2   
\end{tabular}
\caption{Scaling exponents of interface growth, as measured from the results of
numerical simulations, except $\gamma_{\rm calc}$ which was calculated
from $\chi_c$ and $\tau_\parallel$ using (\protect{\ref{sc-par}}) and
(\protect{\ref{sc-opz}}), and $\chi$. The numbers in parenthesis are
errors in the last digit.}
\end{table}


\begin{thebibliography}{99}
\bibitem {tho} The model defined is \cite{high-d} is different than
this definition. The main difference is that there is no well defined
height coordinate for each site of the substrate.
\end{thebibliography}
\end{document}